\documentclass{PoS}
\usepackage{graphicx}

\title{Global fits of the SM parameters}

\ShortTitle{Global fits}

\author{\speaker{Jens Erler}\thanks{On sabbatical leave from IF--UNAM.}\\
        Johannes Gutenberg University Mainz \\
        E-mail: \email{erler@fisica.unam.mx}}

\abstract{I present a global survey of weak mixing angle measurements and other precision experiments
and discuss the issue of correlations of the theoretical uncertainties in precision observables.
Hadronic vacuum polarization effects complicate global fits in various ways
and are also covered alongside some representative fit results.}

\FullConference{7th Annual Conference on Large Hadron Collider Physics - LHCP2019\\
		20-25 May, 2019\\
		Puebla, Mexico}

\begin{document}

\section{Introduction}

I will start by presenting a survey of measurements of the weak mixing angle, $\sin^2\theta_W$, 
as its accurate determination is becoming an ever more global endeavor.
But why are we pushing the precision further and further?
One can compute and measure $\sin^2\theta_W$ and relate it to the $W$ boson mass, $M_W$.
Therefore, one has three ways of obtaining it, yielding a doubly over-constrained system at sub-per mille precision.
And since this system involves relations between couplings and masses of the Standard Model (SM) particles, 
this is {\em the\/} key test of the electroweak symmetry breaking sector.
Moreover, comparisons of measurements at different scales or between different initial or final states
will provide a window to physics beyond the SM that would remain closed if one would have only one kind of determination,
even if that one would be extremely precise.
Thus, a global analysis is important.

There are many approaches to measure $\sin^2\theta_W$.
One is to tune to the $Z$ resonance, where one can measure forward-backward (FB) or left-right (LR) asymmetries
(the latter if one has at least one polarized beam) in $e^+ e^-$ annihilation around the $Z$ boson mass, $M_Z$.
Or one can reverse initial and final states and measure the FB asymmetry in $pp$ or $p\bar p$ Drell-Yan annihilation
in a larger window around $M_Z$.

A very different route is to move to lower energies~\cite{Kumar:2013yoa}, and to consider purely weak processes.
Using neutrinos which do not know about any other interaction,
one can measure the weak mixing angle cleanly in $\nu e$ scattering, 
even though the cross sections are small and therefore statistical uncertainties rather large~\cite{Vilain:1994qy}.
One can move to the deep inelastic scattering (DIS) regime,
where scattering occurs to first approximation off individual quarks, 
as has been done in experiments such as NuTeV~\cite{Zeller:2001hh} at Fermilab.
And very recently the process called Coherent Elastic Neutrino Nucleus Scattering (CE$\nu$NS) as has been observed 
for the first time by the COHERENT Collaboration~\cite{Akimov:2017ade} at Oak Ridge National Laboratory (ORNL).

An alternative strategy to eliminate the electromagnetic interaction is to perform experiments in polarized and therefore
parity-violating electron scattering (PVES)~\cite{Erler:2014fqa},
measuring tiny cross section asymmetries between left-handed and right-handed polarized initial states,
\begin{equation}
A_{LR} = \frac{\sigma_L - \sigma_R}{\sigma_L + \sigma_R}\ .
\end{equation}
Just as for the neutrino case, one may consider a purely leptonic process,
specifically polarized M\o ller scattering, $\vec e^- e^- \to e^- e^-$~\cite{Anthony:2005pm,Benesch:2014bas}.
And again one can scatter deep inelastically (here sometimes called eDIS), 
but there is an important difference to $\nu$DIS mentioned in the previous paragraph.
Because of the small cross sections in neutrino scattering one needs large nuclei, 
which leads to complications from nuclear physics interfering with the unambiguous interpretation of such experiments
(see {\em e.g.}\ Ref.~\cite{Bentz:2009yy}), while in eDIS it is possible to take a target nucleus as small and simple as the deuteron.
Using the 6~GeV CEBAF electron beam, 
this has been done by the PVDIS Collaboration~\cite{Wang:2014bba} at Jefferson Laboratory (JLab).
In fact, polarized eDIS was the key process to establish the SM in 1979~\cite{Prescott:1979dh}, and 
a high-precision measurement will be possible with the SoLID detector~\cite{Souder:2016xcn} at the upgraded CEBAF to 12~GeV.
The PVES analog of CE$\nu$NS on a proton target has been completed very recently by JLab's Qweak 
Collaboration~\cite{Androic:2018kni}, using the nominal 6~GeV CEBAF beam at a lower energy of $E_e = 1.165$~GeV.
This provided the first direct measurement of the weak charge of the proton~\cite{Erler:2003yk}, $Q_W(p)$.
The future P2 experiment~\cite{Becker:2018ggl} at the MESA facility under construction 
at the Johannes Gutenberg University (JGU) at Mainz, will reduce the error in $Q_W(p)$ by a factor of three.
In addition, P2 may also run using a $^{12}$C target which is a very interesting nucleus,
because it is spherical and iso-scalar and has therefore only one nuclear form factor.
Thus, $Q_W(^{12}{\rm C})$ would be easier to interpret,
especially if form factor effects can be constrained by additional run time at larger momentum transfer $Q^2$.
With this at hand, one would be able to disentangle the weak charges of the proton and the neutron,
and consequently the effective vector couplings of the up and down quarks to the $Z$ boson, from PVES alone.

Another new player in this context are isotope ratios in atomic parity violation (APV).
Now, APV in a single isotope is a traditional way to address the neutral-current weak interaction,
and has been studied successfully in heavy alkali atoms~\cite{Wood:1997zq}.
But one faces atomic physics complications, because one needs to understand the atomic structure
in heavy nuclei from sophisticated many-body calculations~\cite{Porsev:2010de,Roberts:2014bka} to a few per mille accuracy.
If, on the other hand, one considers isotope ratios, then most of the atomic physics effects cancel.
The first measurement of APV as a function of isotope number has been achieved by the group of Dima Budker 
at the JGU Mainz~\cite{Antypas:2019raj} just a few months ago,
where the weak charges of Ytterbium showed the expected isotope dependence.

\section{The weak mixing angle}

\begin{figure}[t]
\centering
\includegraphics[width=.73\textwidth]{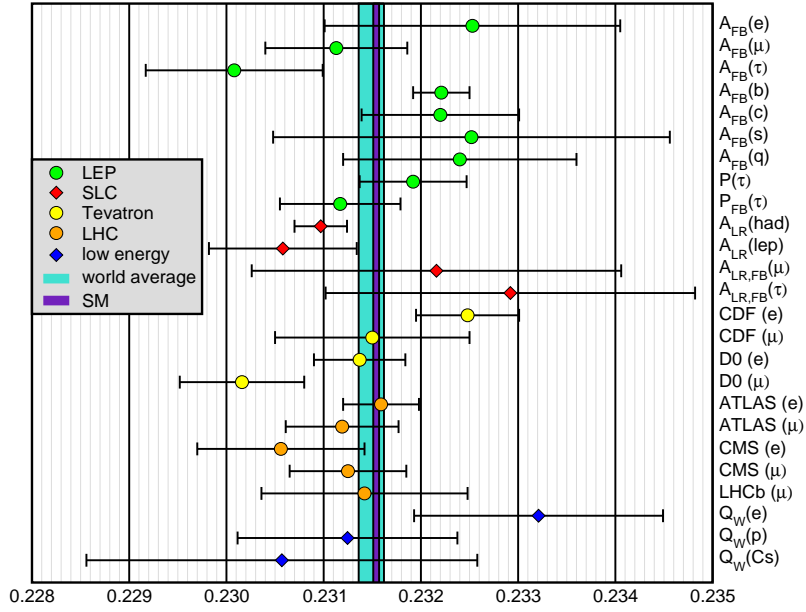}
\caption{Determinations of the {\em effective\/} weak mixing angle, $\sin^2\theta_W$, entering leptonic $Z$ vector couplings.}
\label{sin2thsurvey}
\end{figure}

Figure~\ref{sin2thsurvey} displays all the $\sin^2\theta_W$ determination with sub-percent precision.
The first part are the LEP and SLC measurements in $e^+ e^-$ annihilation near $M_Z$~\cite{ALEPH:2005ab},
yielding the combined result,
\begin{equation}
\sin^2\theta_W ({\rm LEP}) = 0.23153 \pm 0.00016\ .
\end{equation}
Note, that there has been a change in the $\sin^2\theta_W$ extraction from the FB asymmetry for $b\bar b$ final states 
by the LEP Collaborations, as two years ago the two-loop QCD correction necessary to extract the pole asymmetry 
has been obtained with the finite $b$ quark mass dependence~\cite{Bernreuther:2016ccf}, 
reducing the largest LEP discrepancy with the SM by about a quarter of a standard deviation.

Another change affected the extraction from the APV measurement in $^{133}$Cs~\cite{Wood:1997zq},
for which the Stark vector transition polarizability needs to be known.
The latter has been re-measured~\cite{Toh:2019iro} very recently, shifting $|Q_W(^{133}{\rm Cs})|$
which was $1.4~\sigma$ lower than what was expected from the SM much closer to the prediction
(see the lower part of Figure~\ref{sin2thsurvey}).

As for hadron colliders, the leptonic FB asymmetry measurements at the Tevatron combine to the value~\cite{Aaltonen:2018dxj},
\begin{equation}
\sin^2\theta_W ({\rm Tevatron}) = 0.23148 \pm 0.00033\ .
\end{equation}
We averaged the measurements at the LHC by the ALTAS~\cite{ATLAS:2018gqq}, CMS~\cite{Sirunyan:2018swq}, 
and LHCb Collaborations~\cite{Aaij:2015lka},
\begin{equation}
\sin^2\theta_W ({\rm LHC}) = 0.23131 \pm 0.00033\ ,
\end{equation}
by assuming that the smallest theory uncertainty ($\pm 0.00025$ for ATLAS) is common 
to all three detectors~\cite{Erler:2019hds}.
This treatment should be understood as conservative, because the more forward directed kinematics at the LHCb on the one hand,
and the more central kinematics at the general purpose detectors, ATLAS and CMS, on the other hand, 
may in fact provide valuable complementary information on parton distribution functions (PDFs)
which when taken into account is likely to eventually reduce the uncertainty in the combination.
Since rather different aspects of the PDFs are necessary for the extraction of $\sin^2\theta_W$ at $p\bar p$ colliders
compared to those underlying the symmetric $pp$ initial states at the LHC which is based on rapidity distributions,
the corresponding uncertainties can be assumed to be approximately uncorrelated, and we arrive at the world average,
\begin{equation}
\sin^2\theta_W ({\rm world~average}) = 0.23149 \pm 0.00013\ .
\end{equation}
This is in excellent agreement with the result from a global fit to all data,
\begin{equation}
\sin^2\theta_W ({\rm global~fit}) = 0.23153 \pm 0.00004\ .
\end{equation}

\section{The W boson mass}

\begin{figure}[t]
\centering
\includegraphics[width=.6\textwidth]{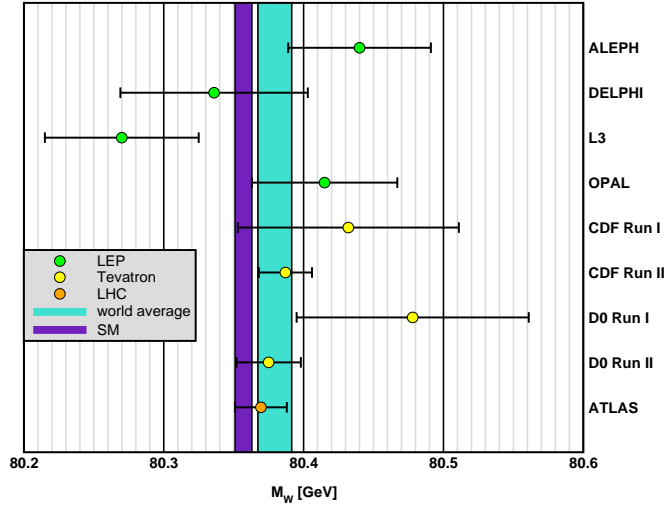}
\caption{Survey of $W$ boson mass measurements.}
\label{mwsurvey}
\end{figure}

Figure~\ref{mwsurvey} shows a similar comparison of $M_W$ determinations.
In contrast to the case of $\sin^2\theta_W$, one observes better mutual agreement among the various measurements 
at LEP~\cite{Schael:2013ita}, the Tevatron~\cite{Aaltonen:2013iut}, and by ATLAS~\cite{Aaboud:2017svj}, but their average,
\begin{equation}
M_W ({\rm world~average}) = 80.379  \pm 0.012~{\rm GeV}\ ,
\end{equation}
is $1.6~\sigma$ higher than the indirect determination of all data excluding the direct measurement results of 
the mass and total width of the $W$ boson,
\begin{equation}
M_W ({\rm indirect}) = 80.357 \pm 0.006~{\rm GeV}\ ,
\end{equation}
and $1.5~\sigma$ higher than the SM prediction,
\begin{equation}
M_W ({\rm global~fit}) = 80.361 \pm 0.005~{\rm GeV}\ .
\end{equation}
The indirect and global fit results for $M_W$ and $\sin^2\theta_W$ account not only for theoretical uncertainties
but also include an estimate and implementation of theoretical correlations~\cite{Erler:2019hds},
which at the achieved level of precision are no longer negligible.

\section{Theoretical uncertainties and the Higgs boson mass}

Indeed, there are various kinds of theory errors entering global fits.
For example, there are the hadronic vacuum polarization and light-by-light contributions obstructing a clean
and unambiguous determination of the anomalous magnetic moment of the muon, $a_\mu$.
There are non-factorizable QCD corrections entering the hadronic $Z$ width~\cite{Freitas:2014hra},
as well as non-resonant corrections to the Breit-Wigner shape of the $Z$ resonance,
but the most important ones are from unknown higher order contributions to the $W$ and $Z$ boson self-energies.
These uncertainties can be estimated by considering the expansion parameters involved.
Including the SM fermion content of three full generations as an enhancement factor, these are
\begin{equation}
\label{eq:enhance1}
\frac{3 \alpha_s}{\pi} \approx 0.116\ , \qquad\qquad 
\frac{8 \alpha}{\pi} \approx 0.020\ ,
\end{equation}
for QCD and QED, and
\begin{equation}
\label{eq:enhance2}
\frac{3 \alpha}{\pi \sin^2\theta_W} \approx 0.032\ , \qquad\qquad
\frac{3 - 6 \sin^2\theta_W + 8 \sin^4\theta_W}{\pi \sin^2\theta_W \cos^2\theta_W}\ \alpha \approx 0.029\ ,
\end{equation}
for the charged and neutral current interactions, respectively.
For the numerical estimates we evaluated the couplings in the $\overline{\rm MS}$ scheme at the $W$ boson mass scale, 
where we have $\alpha^{-1} \approx 128$, $\alpha_s \approx 0.121$, and $\sin^2\theta_W \approx 0.2311$.
Other possible enhancements can arise through the eigenvalue of the quadratic Casimir operator 
in the adjoint representation in QCD, $C_A = 3$, 
and through $\hat m_t^2/M_W^2 \approx 4$ (for the $\overline{\rm MS}$ top quark mass) effects.
We translate these loop factors into uncertainties in the so-called oblique parameters $S = S_Z$, $T$, and 
$U = S_W - S_Z$~\cite{Peskin:1991sw}, which have been originally introduced to parameterize potential new physics contributions to 
electroweak radiative corrections. 
Here, $S_W$ describes the difference of $W$ boson self-energies at the $W$ scale and at very low energies, 
while $T$ refers to the difference of $W$ and $Z$ boson propagator effects at the electroweak scale, and thus violates weak isospin.
Since $S_Z$ is defined in analogy to $S_W$ and receives similar corrections, we rather assume their difference $U$ as a third
{\em uncorrelated\/} uncertainty parameter.
Denoting these uncertainty parameters by $\Delta S_Z$, $\Delta T$ and $\Delta U$, and assuming them to be sufficiently different 
(uncorrelated) {\em induces\/} theory correlations between different observables.
We find $\Delta S_Z = \pm 0.0034$ (which in practice could be added in quadrature to the hadronic vacuum polarization uncertainty
entering the evaluation of the electromagnetic coupling from the Thomson limit to the $Z$ scale),
$\Delta T = \pm 0.0073$ from uncertainties associated with the $tb$ doublet, and $\Delta U = \pm 0.0051$~\cite{Erler:2019hds}.

\begin{figure}[t]
\centering
\includegraphics[width=.71\textwidth]{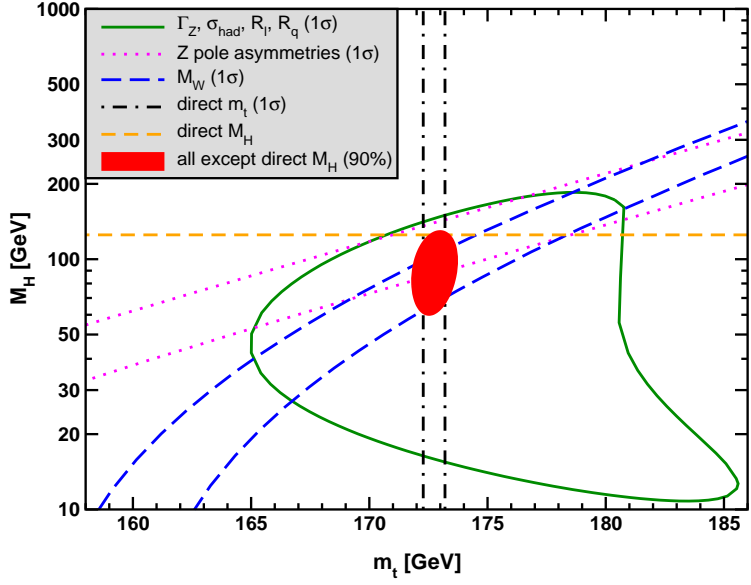}
\caption{Higgs boson mass {\em vs.}\ top quark pole mass for various sets of observables~\cite{Erler:2018pdg}.}
\label{mhmt}
\end{figure}

Figure~\ref{mhmt} reconfirms that $M_W$ is a little deviant from the SM prediction,
while $\sin^2\theta_W$ is in very good agreement.
Indeed, one can determine the top quark mass by means of global fits to all data except for $m_t$ from the Tevatron and the LHC,
and including (excluding) theory uncertainties one obtains\footnote{Note, that due to incremental updates 
some of the numerical results presented here may differ very slightly from those presented at the conference.},
\begin{equation}
m_t ({\rm indirect}) = 176.5 \pm 1.9~(1.8)~{\rm GeV}.
\end{equation}
This represents a $1.8~(1.9)~\sigma$ larger value than the direct measurement~\cite{Erler:2019hds},
\begin{equation}
m_t ({\rm direct}) = 172.90 \pm 0.47~{\rm GeV}. 
\end{equation}
Similarly, one can perform global fits to all data except for the direct $M_H$ constraint from the LHC,
\begin{eqnarray}
M_H ({\rm excluding~theory~uncertainties}) = 90^{+17}_{-15}~{\rm GeV}, \\
M_H ({\rm including~correlated~theory~uncertainties}) = 91^{+18}_{-16}~{\rm GeV},
\end{eqnarray}
showing an only slightly increased central value and uncertainty and a reduced tension (from $1.9~\sigma$ to $1.8~\sigma$) 
with the directly measured value~\cite{Erler:2019hds},
\begin{equation}
\label{eqmh}
M_H ({\rm LHC}) = 125.10 \pm 0.14~{\rm GeV},
\end{equation}
once theory uncertainties are included.

\section{Vacuum polarization in global fits}
I now turn to recent activities in the application of vacuum polarization to global fits.
Firstly, one needs the electromagnetic coupling at the $Z$ peak, $\alpha(M_Z)$, to predict $M_W$ and $\sin^2\theta_W$.
To this end, three groups~\cite{Davier:2017zfy,Jegerlehner:2017zsb,Keshavarzi:2018mgv} have analyzed hadron production data in 
$e^+ e^-$ annihilation, and in some cases also
$\tau$ decay spectral functions which by approximate isospin symmetry yield additional information on the former.
Or one can use perturbation theory for at least part of the calculation, and only rely on data in the hadronic region up to about
2~GeV or so (a quantity we call $\Delta\alpha_{\rm had}^{(3)}(2~{\rm GeV})$), 
and then use the renormalization group equation~\cite{Erler:1998sy,Erler:2017knj}
--- more precisely the anomalous dimension of the photon field ---
which depends on the strong coupling $\alpha_s$, and the charm and bottom quark $\overline{\rm MS}$ masses, 
$\hat{m}_c$ and $\hat{m}_b$.
The results (for the on-shell definition of $\alpha$) are, 
\begin{eqnarray}
\label{alphaMZ1}
\alpha^{-1} (M_Z) = 128.947 \pm 0.012~\cite{Davier:2017zfy}, &\\
\alpha^{-1} (M_Z) = 128.958 \pm 0.016~\cite{Jegerlehner:2017zsb}, &\\
\alpha^{-1} (M_Z) = 128.946 \pm 0.015~\cite{Keshavarzi:2018mgv}, &\\
\alpha^{-1} (M_Z) = 128.949 \pm 0.010~\cite{Erler:2017knj}, &
\label{alphaMZ4}
\end{eqnarray}
where the various authors use slightly different input values for $\alpha_s(M_Z)$, 
but this amounts to differences below the level of 0.004 in $\alpha^{-1} (M_Z)$.

Interestingly, the data used for the hadronic part also enter other observables present in global electroweak fits,
which induces another source of uncertainty correlation.
For example, these data play a crucial role in the evaluation of the SM prediction of $a_\mu$,
where they enter first at the two-loop level and generate a correlation with $\alpha(M_Z)$,
and both are in turn anti-correlated with the three-loop vacuum polarization contribution to $a_\mu$.
Finally, there is a correlation with the quantity $\sin^2\theta_W(0) - \sin^2\theta_W(M_Z)$.

Because the muon mass scale is rather low, most of the evaluation of the hadronic vacuum polarization contribution to $a_\mu$
needs to be based on data.
However, there is a fraction that can be computed perturbatively. 
In particular, the heavy quark contributions are fully accessible in perturbation theory~\cite{Erler:2000nx},
which for the charm and bottom contributions yields,
\begin{eqnarray}
\label{amuc}
&&a_\mu^c ({\rm vacuum~polarization; PQCD}) = (14.6 \pm 0.5_{\rm PQCD} \pm 0.2_{\hat{m}_c} \pm 0.1_{\alpha_s}) 10^{-10}, \\
&&a_\mu^b ({\rm vacuum~polarization; PQCD}) =  0.3 \times 10^{-10},
\end{eqnarray}
where the errors in Equation~(\ref{amuc}) are from the truncation of the perturbative series at order $\alpha_s^2$, 
and the parametric uncertainties in $\hat{m}_c(\hat{m}_c)$ (taken from Ref.~\cite{Erler:2016atg}) and from $\alpha_s$.
Equation~(\ref{amuc}) is in excellent agreement 
with the very recent result obtained from a lattice gauge theory simulation~\cite{Gerardin:2019rua},
\begin{equation}
a_\mu^c ({\rm vacuum~polarization; lattice}) = (14.66 \pm 0.45_{\rm stat.} \pm 0.06_{\rm syst.}) 10^{-10},
\end{equation}
and is of similar precision.  
Since the bottom quark introduces a rather different scale, its effect is much harder to study on the lattice, 
while the perturbative method is even more reliable here, yielding a negligible error.
It should also be remarked that Ref.~\cite{Gerardin:2019rua} finds a rather large total hadronic vacuum polarization effect,
so that if confirmed, there would cease to be a conflict between the measurement of $a_\mu$~\cite{Bennett:2006fi} and the SM,
which currently amounts to more than $3~\sigma$.
But then there would be a new discrepancy between the dispersive and lattice approaches to vacuum polarization.

\begin{figure}[t]
\centering
\includegraphics[width=.89\textwidth]{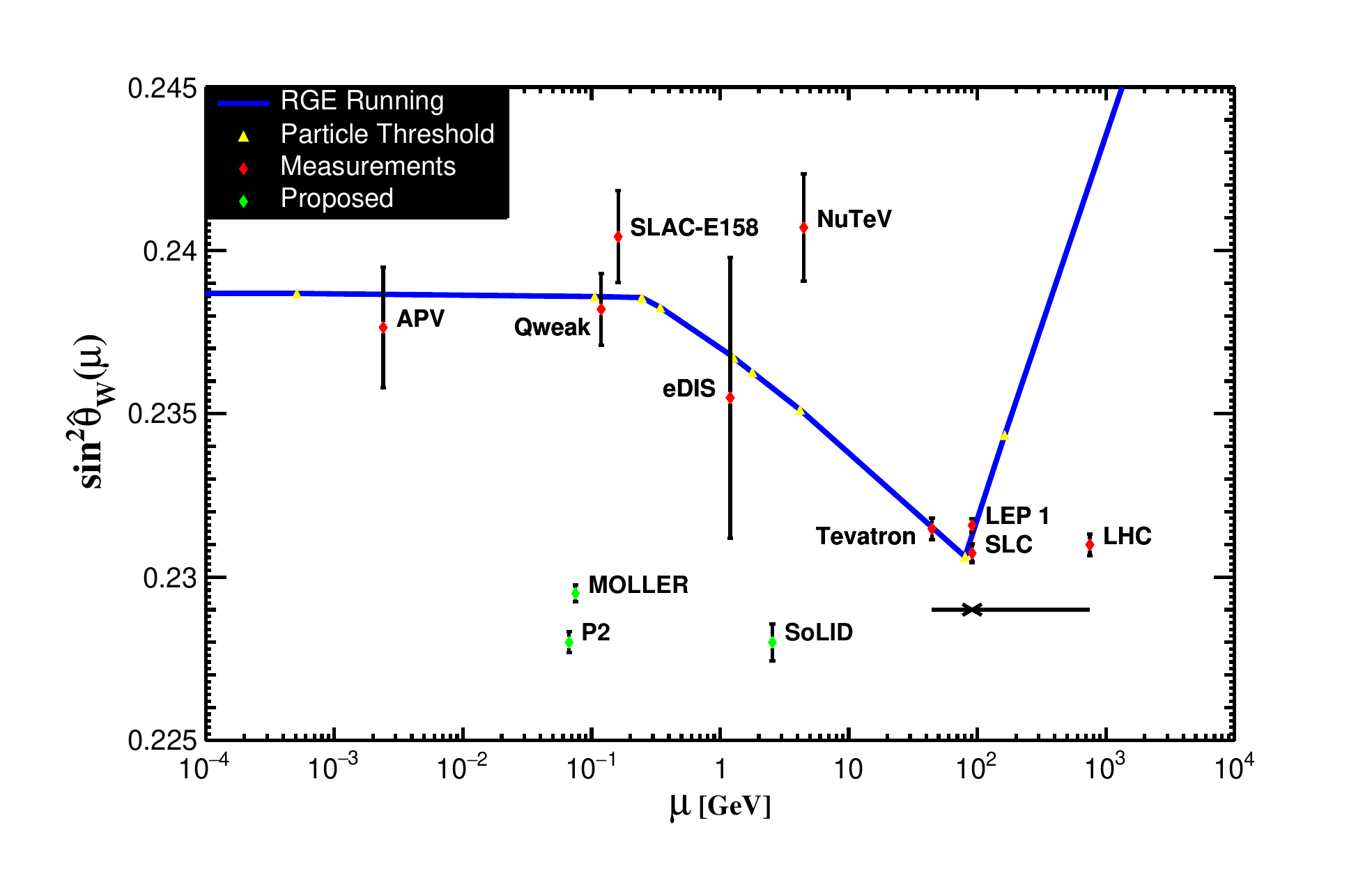}
\caption{Renormalization group evolution (running) of the weak mixing angle (updated from Ref.~\cite{Erler:2017knj}).}
\label{runnings2w}
\end{figure}

An important quantity where vacuum polarization enters is $\sin^2\theta_W(0)$.
It is needed for many low-energy electroweak observables, and it can be seen from Figure~\ref{runnings2w} 
that future experiments in low momentum transfer PVES will be at the precision level of the LEP and SLC measurements.
As mentioned earlier, the interpretation of NuTeV and other $\nu$DIS experiments suffers from poorly understood and
potentially large nuclear effects, and should probably be best removed from this plot which ought to illustrate 
the experimental verification of the running of $\sin^2\theta_W$.

There are basically three parts in the computation of $\sin^2\theta_W(0)$.
The largest piece can be obtained by solving the coupled system of differential equations for the $\overline{\rm MS}$ quantities
$\hat\alpha$ and $\sin^2\hat\theta_W$,
\begin{eqnarray}
&&\mu^2 \frac{d\hat\alpha}{d\mu^2} = \frac{\hat\alpha^2}{\pi} \left[\frac{1}{24} \sum_i K_i \gamma_i Q_i^2 + 
\sigma\left(\sum_q Q_q \right)^2 \right], \\
&&\mu^{2}\frac{d\hat{v}_{f}}{d\mu^{2}} = \frac{\hat\alpha Q_f}{24\pi}
\left[ \sum_i K_i \gamma_i \hat{v}_i Q_i + 12 \sigma \left( \sum_q Q_q \right) 
\left( \sum_q \hat{v}_q \right)\right],
\end{eqnarray}
where $\hat v_f = T_f - 2 Q_f \sin^2\hat\theta_W$ is the $Z$ boson vector coupling to fermion $f$,
and $T_f$ is the third component of weak isospin of fermion $f$.
The sums are over all active particles in the relevant energy range.
The $Q_i$ are the electric charges, while the $\gamma_i$ are constants depending on the field type~\cite{Erler:2004in}.
For quarks, $K_i$ is proportional to the color factor $N_C = 3$ and contains QCD corrections up to and including 
${\cal O}(\alpha_s^4)$~\cite{Erler:2017knj}, while for both leptons and bosons $K_i = N_C = 1$.
The terms involving $\sigma$ arise due to QCD annihilation (singlet) diagrams with purely gluonic sub-diagrams.
We can relate the renormalization group equation for $\hat{\alpha}$ to that for $\sin^2\hat\theta_W$ since both, 
the $\gamma Z$ mixing tensor $\Pi_{\gamma Z}$ and the photon vacuum polarization function $\Pi_{\gamma\gamma}$,
are pure vector-current correlators. 
As already mentioned, the remaining non-perturbative part needs the same kind of data that entered the calculations 
of $\alpha(M_Z)$ to obtain Equations~(\ref{alphaMZ1})--(\ref{alphaMZ4}).
This part needs to be subdivided into two pieces because the vector couplings of the $Z$ boson differ from the electric charges,
implying that there is a piece that is not directly related to $\alpha(M_Z)$ and necessitating a study of the effect and
uncertainty associated with the corresponding flavor separation into up- and down-type quarks.
One also needs to estimate the uncertainties from the singlet piece and from isospin breaking effects.
The overall uncertainty is negligible compared
to any upcoming low-energy determination of $\sin^2\theta_W$ in the foreseeable future.
It is interesting that if one were to compute $\sin^2\theta_W(0)$ directly rather than relative to $\sin^2\theta_W(M_Z)$,
the hadronic vacuum polarization data would enter twice, and their errors would {\em add\/}.
This is because 
\begin{equation}
\sin^2\theta_W(0) = \hat\kappa(0) \sin^2\theta_W(M_Z),
\end{equation}
and the correction factor $\hat\kappa(0)$ increases with larger $e^+e^-$ hadronic cross sections,
and $\sin^2\theta_W(M_Z)$ itself is proportional to $\alpha(M_Z)$ which can be seen by recalling that 
\begin{equation}
M_Z^2 \propto g_Z^2 (M_Z) v^2 \propto \frac{\alpha(M_Z)}{\sin^2\theta_W(M_Z) \cos^2\theta_W(M_Z)} G_F^{-1},
\end{equation}
and solving for $\sin^2\theta_W(M_Z)$ ($v$ is the Higgs boson vacuum expectation value and $G_F$ is the Fermi constant).

\begin{figure}[t]
\centering
\includegraphics[width=.6\textwidth]{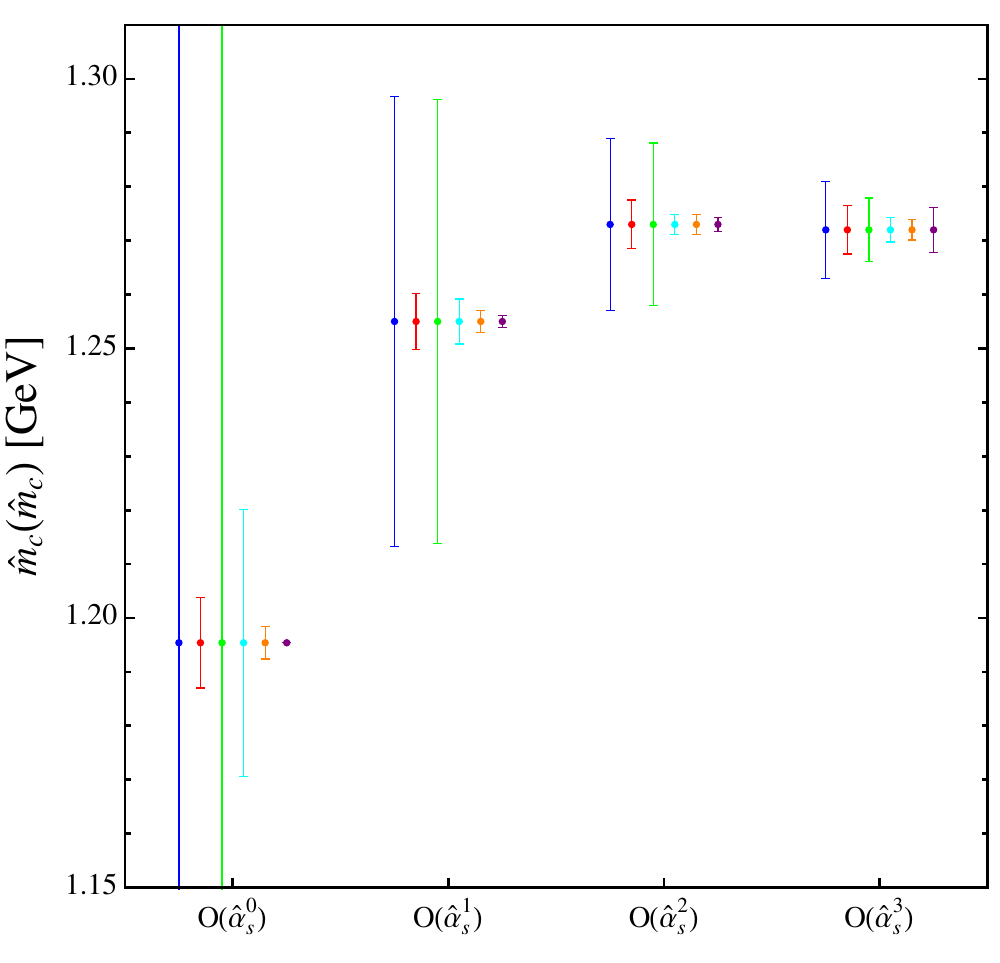}
\caption{Error budget for the sum rule analysis for $\hat{m}_c(\hat{m}_c)$ as a function of loop order~\cite{Erler:2016atg}.}
\label{mcplot}
\end{figure}

The final application of vacuum polarization are charm and bottom quark mass determinations.
If one employs as input quantities only the electronic decay widths of the narrow resonances, {\em e.g.}, the $J/\psi$ 
and the $\psi(2S)$ in the case of charm, and compares two different moments of the relevant vacuum polarization function,
one obtains simultaneous information on the quark mass and the less well-known continuum contribution.
The constraint on the latter can then be compared with the corresponding moments of the experimental determinations 
of electro-production of open charm or bottom.
This results in over-constrained systems, where any residual difference can be taken as an error
estimate~\cite{Erler:2016atg,Erler:2002bu} of non-perturbative effects, which are supposedly small but possibly not entirely negligible. 
This strategy has been applied to $\hat{m}_c$ resulting in the precision determination~\cite{Erler:2016atg},
\begin{equation}
\hat{m}_c(\hat{m}_c) = 1272 \pm 8 + 2616 [\alpha_s(M_Z) - 0.1182]~{\rm MeV},
\end{equation}
where the central value is in very good agreement with recent lattice results~\cite{Aoki:2019cca} and of similar precision.
The breakdown of the error in this approach is detailed in Figure~\ref{mcplot}.

\begin{table}[t]
\centering
\begin{tabular}{|c|c|rrrrrr|}
\hline
$M_Z$ & $91.1884 \pm 0.0020$~GeV & 1.00 &$-0.06$ & 0.00 & 0.00 & 0.02 & 0.03 \\
$\hat{m}_t(\hat{m}_t)$ & $163.28\pm 0.44$~GeV & $-0.06$ & 1.00 & 0.00 & $-0.13$ & $-0.28$ & 0.03 \\
$\hat{m}_b(\hat{m}_b)$ & $4.180\pm 0.021$~GeV & 0.00 & 0.00 & 1.00 & 0.00 & 0.00 & 0.00 \\
$\hat{m}_c(\hat{m}_c)$ & $1.275 \pm 0.009$~GeV & 0.00 & $-0.13$ & 0.00 & 1.00 & 0.45 & 0.00 \\
$\alpha_s(M_Z)$ & $0.1187 \pm 0.0016$ & 0.02 &$-0.28$ & 0.00 & 0.45 & 1.00 & $-0.02$ \\
$\Delta\alpha_{\rm had}^{(3)}(2~{\rm GeV} $ & $0.00590\pm 0.00005$ & 
0.03 & 0.03 & 0.00 & 0.00 & $-0.02$ & 1.00 \\ 
\hline
\end{tabular}
\caption{Standard global fit~\cite{Erler:2018pdg},
where the correlation with $M_H$ is Equation~(\ref{eqmh}) is negligible.}
\label{tab:SMresults}
\end{table}

\section{Results and conclusions}
Table~\ref{tab:SMresults} together with Equation~(\ref{eqmh}) summarizes the SM global fit~\cite{Erler:2018pdg}.
Two simple examples are to suffice to illustrate the application of global fits to constraints on physics beyond the SM.

One is the $\rho_0$ fit, in which one assumes that the new physics is mainly affecting the $\rho$ parameter,
which is a measure of the neutral-to-charged current interaction strengths.
Various quantities constrain $\rho_0$ and there are different ways to interpret it.
For example, any electroweak doublet with a non-trivial mass splitting,
$\Delta m_i^2 \geq (m_1 - m_2)^2$, contributes to $\rho_0$ positive definitely,
\begin{equation}
\Delta \rho_0 =  \sum_i \frac{G_F N_C}{8\sqrt{2} \pi^2} \Delta m_i^2\ .
\end{equation}
This equation might appear to suggest that there is no decoupling, so that even a doublet with grand unification 
or Planck scale masses but electroweak size splitting may give observable effects in experiments at much lower energies.
However, this is not the case, as there is a see-saw type suppression of $\Delta m_i^2$ in any given model.
Another way of saying this is to recall that the leading contribution to $\rho_0$ in the SM effective field theory 
is a combination of dimension six operators, so that these effects are suppressed
on dimensional grounds by at least two powers of the scale of new physics.
The global fit yields~\cite{Erler:2018pdg},
\begin{equation}
\rho_0 =  1.00039 \pm 0.00019,
\end{equation}
which is 2~$\sigma$ higher than the SM value, $\rho_0 \equiv 1$.
This is another manifestation of the tension in the $W$~boson mass discussed earlier.   
It is amusing to point out that at face value, one even finds a non-trivial 95\%~CL {\em lower\/} bound
on the sum of all such mass splittings,
\begin{equation}
(16~{\rm GeV})^2 \leq \sum_i \frac{N_C}{3} \Delta m_i^2 \leq (48~{\rm GeV})^2 \hspace{36pt} (90\%~{\rm CL}).
\end{equation}
This strongly disfavors, for example, zero hypercharge, $Y = 0$, Higgs triplets for which $\rho_0 < 1$.
On the other hand, a Higgs triplet with $|Y| = 1$ is consistent with the data provided its vacuum expectation value
is around 1\% of that of the SM doublet.

\begin{figure}[t]
\centering
\includegraphics[width=.7\textwidth]{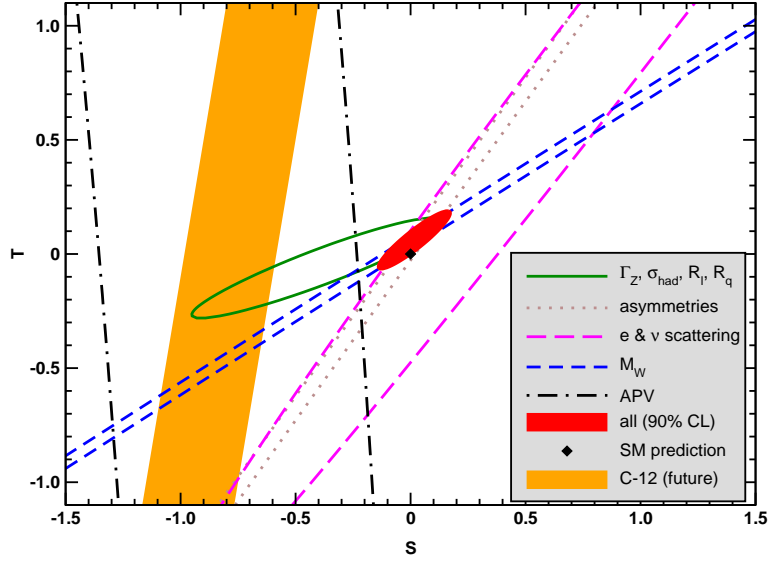}
\caption{$T$ {\em versus\/} $S$ for various data sets (modified from Ref.~\cite{Erler:2018pdg}).  
For illustration, I also show the impact that the $^{12}$C PVES measurement would have 
if it could be performed with a relative error of $0.3\%$, dominated by the polarization uncertainty
(within the SM another measurement of $\sin^2\theta_W$ to which the polarization asymmetry is directly proportional).
As can be seen, this yields a different slope in the $ST$-plane.}
\label{STplot}
\end{figure}

The other example is the fit result~\cite{Erler:2018pdg} for the $S$ and $T$ parameters~\cite{Peskin:1991sw}, 
\begin{eqnarray}
&& S = 0.02 \pm 0.07, \\
&& T = 0.06 \pm 0.06, 
\end{eqnarray}
with a correlation of 81\%. 
This fit is illustrated in Figure~\ref{STplot} and fixes $U = 0$, as $U$ is expected to be suppressed by two additional factors
of the new physics scale compared to $S$ and $T$~\cite{Grinstein:1991cd}.
It is remarkable that the $\chi^2$ value at the minimum of the fit drops by 4.2~units if these two extra degrees
of freedom are allowed.
This kind of improvement is non-trivial and quite rare, and is again related to $M_W$.
One can interpret the $S$ and $T$ parameters in a variety of new physics models,
if one assumes that non-oblique effects are absent or small.
For example, the mass of the lightest Kaluza-Klein state~\cite{Carena:2006bn} in warped extra dimensions~\cite{Randall:1999ee}
should satisfy the bound $M_{KK} \gtrsim 3.2$~TeV~\cite{Erler:2018pdg}, while the lightest vector state 
in minimal composite Higgs models~\cite{Pich:2013fea} is bound by $M_V \gtrsim 4$~TeV~\cite{Erler:2018pdg}.

To conclude, both, the LHC and low-energy measurements are approaching LEP and SLC precision in $\sin^2\theta_W$.
There are new players represented by COHERENT~\cite{Akimov:2017ade}, Qweak~\cite{Androic:2018kni}, 
and APV isotope ratios~\cite{Antypas:2019raj}, which are not quite as precise yet, but they are first measurements
with great prospects of improvement.
As for today, with the lower precision with which the results of these first measurements came in,
it is more interesting to assume the validity of the SM, and use them to constrain neutron skins
(the difference of the neutron and proton radii in nuclei), or more generally form factor effects.

For ultra-high precision tests, not only theoretical uncertainties from unknown higher orders need to be estimated
and included, but also their correlations across various observables, even though this is a difficult task.
As examples, I discussed uncertainties in the $W$ and $Z$ boson self-energies due to unknown higher orders,
as well as hadronic vacuum polarization effects and their uncertainties which enter correlated in an increasing number of quantities.

\section*{Acknowledgements}
\noindent
This work is supported by CONACyT (M\'exico) project 252167--F,
the German--Mexican research collaboration grant SP 778/4--1 (DFG) and 278017 (CONACyT),
and by PASPA (DGAPA--UNAM).
I gratefully acknowledge the outstanding hospitality and support offered by 
the Helmholtz-Institute Mainz,
the Theoretical High Energy Physics (THEP) group at the Institute of Physics in Mainz,
the Mainz Institute for Theoretical Physics (MITP), 
and the PRISMA$^+$ cluster of excellence, where a significant part of this work has been carried out. 
Finally, I thank Rodolfo Ferro-Hern\'andez for updating Figure~\ref{runnings2w}, and
I am indebted to Werner Bernreuther and Long Chen for a dedicated update and a private communication 
w.r.t.\ the result of Ref.~\cite{Bernreuther:2016ccf} to incorporate it into the program GAPP.

\end{document}